\documentclass[12pt]{iopart}

\usepackage{graphicx}
\usepackage{bm}
\usepackage{times}
\usepackage{amsmath,amssymb}
\usepackage{amsthm}
\usepackage{enumerate}
\usepackage{multirow}
\usepackage{qcircuit}
\usepackage[numbers,sort&compress]{natbib}

\usepackage[colorlinks=true,linkcolor=blue,citecolor=red, linktocpage=true,breaklinks=true]{hyperref}

\newtheorem{theorem}{Theorem}

\begin{document}

\title{Geometric representations of braid and Yang-Baxter gates}

\author{Kun Zhang$^{1, 2, 3}$, Kun Hao$^{3, 4, \dag}$, Kwangmin Yu$^{5}$, Vladimir Korepin$^{6}$, Wen-Li Yang$^{2,3,4}$}

\address{$^1$ School of Physics, Northwest University, Xi’an 710127, China}
\address{$^2$ Shaanxi Key Laboratory for Theoretical Physics Frontiers, Xi'an 710127, China}
\address{$^3$ Peng Huanwu Center for Fundamental Theory, Xi'an 710127, China}
\address{$^4$ Institute of Modern Physics, Northwest University, Xi'an 710127, China}
\address{$^5$ Computational Science Initiative, Brookhaven National Laboratory, Upton, New York 11973, USA}
\address{$^6$ C.N. Yang Institute for Theoretical Physics, Stony Brook University, Stony Brook, New York 11794, USA}

\ead{haoke72@163.com}

\vspace{10pt}
\begin{indented}
\item[]June 2024
\end{indented}

\begin{abstract}
Brick-wall circuits composed of the Yang-Baxter gates are integrable. It becomes an important tool to study the quantum many-body system out of equilibrium. To put the Yang-Baxter gate on quantum computers, it has to be decomposed into the native gates of quantum computers. It is favorable to apply the least number of native two-qubit gates to construct the Yang-Baxter gate. We study the geometric representations of all X-type braid gates and their corresponding Yang-Baxter gates via the Yang-Baxterization. We find that the braid and Yang-Baxter gates can only exist on certain edges and faces of the two-qubit tetrahedron. We identify the parameters by which the braid and Yang-Baxter gates are the Clifford gate, the matchgate, and the dual-unitary gate. The geometric representations provide the optimal decompositions of the braid and Yang-Baxter gates in terms of other two-qubit gates. We also find that the entangling powers of the Yang-Baxter gates are determined by the spectral parameters. Our results provide the necessary conditions to construct the braid and Yang-Baxter gates on quantum computers. 

\end{abstract}

%
%

\section{\label{sec:intro} Introduction}

Quantum computers can efficiently solve problems that are unrealistic to be solved by classical computers \cite{nielsenQuantumComputationQuantum2010}. Simulating many-body quantum systems is a major application of quantum computers in the current NISQ era \cite{preskill2018quantum}. The most popular model for quantum computation is the quantum circuit model. It is the quantum version of the classical reversible computation model. The quantum circuit model is composed of the initial state, unitary evolution realized by certain single- and two-qubit gates, and the final state measurements. Arbitrary unitary evolution can be approximately constructed from certain single- and two-qubit gates, known as the universal gate set \cite{divincenzo1995two,BBCDMSSSW95}. 

Although very few many-body systems can be exactly solved, these models lay the theoretical foundations for our understanding of the many-body system \cite{sutherland2004beautiful,baxter2016exactly}. One important category is called the Yang-Baxter integrable models, which are solved based on the Yang-Baxter equation \cite{yang1967some,yang1968s,baxter1972partition}. The systematic method solving the Yang-Baxter integrable models is called the algebraic Bethe ansatz or the quantum inverse scattering method \cite{takhtadzhan1979quantum,korepin1997quantum}. See a recent review on the Yang-Baxter integrable models and their applications \cite{batchelor2016yang}.

The essential ingredient to solve the Yang-Baxter integrable models is the $R$ matrix, the solution of the Yang-Baxter equation. In the context of quantum computation, the unitary $R$ matrix can work as a quantum gate, called the Yang-Baxter gate \cite{dye2003unitary}. In the last twenty years, the interdisciplinary studies between the quantum integrable system and quantum computation, especially based on the quantum circuit model, were developed in different directions, in either of which the Yang-Baxter gate plays fundamental roles. 

Unitary solutions of parameter independent Yang-Baxter equation, also called the constant Yang-Baxter equation, give representations of braid group \cite{kauffman2001knots}. Here the Yang-Baxter gate is also called the braid gate. It can characterize the low-dimensional topology but also can be viewed as a quantum gate generating the quantum entanglement \cite{kauffman2002quantum,kauffman2004braiding}. Therefore, the manipulation of quantum entanglement via the braid gate can be studied from the viewpoint of topology \cite{alagic2016yang,kauffman2019topological,quinta2018classifying,padmanabhan2021local}. Then various quantum information protocols, such as quantum teleportation and entanglement swapping, can be understood from the viewpoint of topological entanglement \cite{zhang2006teleportation,chen2007braiding,zhang2016teleportation,zhang2017quantum}. 

In another research direction, the Yang-Baxter gate with the spectral parameter arises from the quantum simulation (based on the circuit model) of integrable models. It is surprising that the simulation circuit obtained from decomposing the evolution operator into two-qubit gates, called the Trotterization, is still integrable \cite{vanicat2018integrable}. The spectral parameter of the Yang-Baxter gate is interpreted as the Trotter step. Taking arbitrary values of the spectral parameter, the brick-wall circuit composed of the Yang-Baxter gates is always integrable \cite{miao2024floquet}. Besides the Trotterization method, the integrable circuits are also developed from the cellular automata theory, where the medium-range integrable circuits are also constructed \cite{pozsgay2021yang,gombor2021integrable,gombor2022superintegrable,gombor2024integrable}. The integrable circuit preserves many nice properties of the integrable model. For example, the spectrum has multi-particle bound state solutions \cite{aleiner2021bethe}; the circuit has an infinite number of conserved charges \cite{maruyoshi2023conserved}; the correlation functions can be exactly computed \cite{claeys2022correlations,hutsalyuk2024exact}; the circuit has a ballistic spin transport \cite{ljubotina2019ballistic,zadnik2024quantum}; the circuit exists strong zero modes at the boundary \cite{vernier2024strong}. In experiments, the integrable circuit can not only benchmark quantum computers but also provides a testbed to study the integrable-breaking dynamics \cite{morvan2022formation,kim2023evidence,keenan2023evidence,shtanko2023uncovering}.


The Yang-Baxter equation is also called the factorization equation in some literature, which describes the consistent relation for decomposing the three-body scattering matrix into the two-body ones. In the context of the quantum circuit model, the quantum computation is realized by specific single- and two-qubit gates, namely the universal gate set. Therefore, we may view the quantum circuit as a ``factorized'' evolution. In the sense of factorization, the integrable circuit may suggest a deep understanding of the quantum circuit model \cite{zhang2012quantum,zhang2013integrable,banchi2017quantum}.

Despite the vital importance of the Yang-Baxter gate, its completed characterizations are missing. For example, which two-qubit gates can be converted into the Yang-Baxter gates through single-qubit gates? Which Yang-Baxter gates can be more easily realized on quantum computers (from the native two-qubit gates)? How to decompose the Yang-Baxter gate with the minimal applications of the native gates of quantum computers? We aim to answer the above questions by studying the geometric representation of braid and Yang-Baxter gates. Here the geometric representation refers to mapping the two-qubit gates on a three-dimensional tetrahedron, where the two-qubit gates on the same point are locally equivalent (can be converted by single-qubit gates) \cite{zhang2003geometric}. Then we give the optimal decompositions (with respect to the number of CNOT) of the braid and Yang-Baxter gates. In our study, we consider all the X-type braid gates and the Yang-Baxter gates obtained from braid gates through the Yang-Baxterization \cite{jonesBaxterization1990,chengYangBaxterizationBraidGroup1991,geExplicitTrigonometricYangbaxterization1991}. There are other special types of two-qubit gates, such as the Clifford gate \cite{gottesmanStabilizerCodesQuantum1997,gottesmanTheoryFaulttolerantQuantum1998}, the matchgate \cite{valiantQuantumComputersThat2001,jozsaMatchgatesClassicalSimulation2008}, and the dual unitary gate \cite{bertiniExactCorrelationFunctions2019}. The circuits composed of the Clifford gate or the matchgate (acting on the nearest-neighbor qubits) can be classically simulated. The correlation function of the brick-wall circuits composed of the dual unitary gate can be exactly computed. We clarify the conditions under which the braid and Yang-Baxter gates are the Clifford gates, the matchgates, and the dual unitary gates. We calculate their entangling powers, which have been applied to the study of brick-wall circuits \cite{hahn2024absence}.  

Our paper is organized as follows. In Sec. \ref{sec:two_qubit}, we review basic theories about the two-qubit gate, as well as its geometric representation. We present our results, namely the geometric representations of the braid gate and the Yang-Baxter gate in Secs. \ref{sec:Geo_braid} and \ref{sec:YBG} respectively. Sec. \ref{sec:conclusion} is the conclusion. \ref{App:A} gives the proof of Theorem \ref{theorem_real_x}. \ref{App:B} shows the matrix expressions of the Yang-Baxter gates obtained from the Yang-Baxterization of the braid gates.

\section{\label{sec:two_qubit}Tetrahedron of Two-qubit gates}

\subsection{\label{subsec:nonlocal_entangling_power}Nonlocal parameters and entangling power}

Two-qubit gates $U$ are elements of the $U(4)$ group. Since the global phase has no significance in quantum computation, we concentrate on the two-qubit gates belonging to the $SU(4)$ group. The nontrivial two-qubit gates are those that cannot be constructed from the single-qubit gates, namely $U\in SU(4)\backslash SU(2)\otimes SU(2)$. Then one can apply the Cartan decomposition to $U\in SU(4)\backslash SU(2)\otimes SU(2)$, which gives \cite{krausOptimalCreationEntanglement2001a}
\begin{equation}
\label{eq:U_decomposition}
    U = (V_1\otimes V_2)U_\text{core}(\vec a)(V_3\otimes V_4),
\end{equation}
with the core of the two-qubit gate
\begin{equation}
\label{eq:def_U_core}
    U_\text{core}(\vec a) = \exp\left(\frac i 2\left(a_1(\sigma_x\otimes \sigma_x) + a_2(\sigma_y\otimes \sigma_y) + a_3(\sigma_z\otimes \sigma_z)\right)\right).
\end{equation}
Here $V_k\in SU(2)$ ($k=1,2,3,4$) are the single-qubit gates and $\sigma_{x,y,z}$ are Pauli matrices. Therefore, only $\vec a = (a_1,a_2,a_3)$, called the nonlocal parameters, characterizes the intrinsic properties of two-qubit gates. For simplicity, we use
\begin{equation}
    [U] = [a_1,a_2,a_3],
\end{equation}
to denote the nonlocal parameters of $U$. We say that two-qubit gates are locally equivalent if these two-qubit gates have the same nonlocal parameters. In other words, two-qubit gates with the same nonlocal parameters only differ by some single-qubit gates. In the two-qubit computational basis $\{|00\rangle,|01\rangle,|10\rangle,|11\rangle\}$, the core gate has the expression
\begin{equation}
\small
    U_\text{core}(\vec a) = \begin{pmatrix}
        e^{i\frac{a_3}{2}}\cos\left(\dfrac{a_1-a_2}{2}\right) & 0 & 0 & i e^{i\frac{a_3}{2}}\sin\left(\dfrac{a_1-a_2}{2}\right) \\
        0 & e^{-i\frac{a_3}{2}}\cos\left(\dfrac{a_1+a_2}{2}\right) & i e^{-i\frac{a_3}{2}}\sin\left(\dfrac{a_1+a_2}{2}\right) & 0 \\
        0 & i e^{-i\frac{a_3}{2}}\sin\left(\dfrac{a_1+a_2}{2}\right) & e^{-i\frac{a_3}{2}}\cos\left(\dfrac{a_1+a_2}{2}\right) & 0 \\
        i e^{i\frac{a_3}{2}}\sin\left(\dfrac{a_1-a_2}{2}\right) & 0 & 0 & e^{i\frac{a_3}{2}}\cos\left(\dfrac{a_1-a_2}{2}\right) \\
    \end{pmatrix}.
\end{equation}
The half entries are zero. It is an example of an X-type two-qubit gate since the nonzero entries form an ``X'' structure. Note that the $R$ matrix of the eight-vertex model also has the ``X'' structure.

Quantities associated with the two-qubit gate, which are invariant under the action of single-qubit gates, are called local invariants. It is expected that the local invariants are given by $[a_1,a_2,a_3]$. It has been shown that the local invariants of the two-qubit gates are functions of $\tr(\Lambda)$ and $\tr(\Lambda^2)$, where $\Lambda$ is a diagonal matrix given by $\Lambda = \left\{e^{i(-a_1+a_2+a_3)},e^{i(a_1-a_2+a_3)},e^{i(a_1+a_2-a_3)},e^{-i(a_1+a_2+a_3)}\right\}$ \cite{makhlinNonlocalPropertiesTwoqubit2002}. The four diagonal terms in $\Lambda$ are eigenvalues of $U_Q^TU_Q$. Here $U_Q$ is obtained from the two-qubit core $U_\text{core}(\vec a)$ defined in Eq. (\ref{eq:def_U_core}) by the transformation $U_Q = Q^TU_\text{core}(\vec a) Q$. Here $Q$ is a transformation from the two-qubit computational basis to the Bell basis. More details can be found in \cite{makhlinNonlocalPropertiesTwoqubit2002}.


Two-qubit gates with different nonlocal parameters may have different abilities to generate the entanglement. For example, the CNOT gate given by
\begin{equation}
    \text{CNOT} = |0\rangle\langle 0|\otimes 1\!\!1_2 + |1\rangle\langle 1|\otimes\sigma_x,
\end{equation}
has the nonlocal parameters $[\text{CNOT}] = [\pi/2,0,0]$, which can generate Bell states from the product states (with the help of single-qubit gates). However, the SWAP gate defined by
\begin{equation}
    \text{SWAP} = \frac 1 2 \left(1\!\!1_2+\sigma_x\otimes\sigma_x + \sigma_y\otimes\sigma_y + \sigma_z\otimes\sigma_z\right)
\end{equation}
has $[\text{SWAP}] = [\pi/2,\pi/2,\pi/2]$, which only permutes between two qubits, and can not create any entanglement from the product states. For the XY interaction qubit, it is natural to give the iSWAP gate \cite{tanamoto2009efficient}
\begin{equation}
\label{eq:def_iswap}
    \text{iSWAP} = \frac 1 2 \left(1\!\!1_2+i\sigma_x\otimes\sigma_x + i\sigma_y\otimes\sigma_y + \sigma_z\otimes\sigma_z\right),
\end{equation}
which has $[\text{iSWAP}] = [\pi/2,\pi/2,0]$. Therefore, it can generate entangled states from the product states. 

To quantify the entanglement-generation ability of two-qubit gates, consider
\cite{zanardiEntanglingPowerQuantum2000}
\begin{equation}
    e_p(U) = \overline{E(U|\psi_1\rangle\otimes|\psi_2\rangle)}_{|\psi_1\rangle\otimes|\psi_2\rangle},
\end{equation}
with the linear entropy
\begin{equation}
\label{def:linear_entropy}
    E(|\Psi\rangle) = 1 - \tr_1\rho^2,
\end{equation}
and $\rho = \tr_2|\Psi\rangle\langle\Psi|$. The overline means taking the average over all product states $|\psi_1\rangle\otimes|\psi_2\rangle$ with the uniform distribution. We call $e_p(U)$ as the entangling power of two-qubit gates. Because of the simple decomposition given in Eq. (\ref{eq:U_decomposition}), one can find that the entangling power of the two-qubit gate is \cite{balakrishnanEntanglingPowerLocal2010}
\begin{equation}
\label{def:entangling_power}
    e_p(U) = \frac 2 9 \left(1 - \frac{1}{16}\left|\tr(\Lambda)\right|^2\right),
\end{equation}
with the local invariants of the two-qubit gate
\begin{equation}
    \left|\tr(\Lambda)\right|^2 = 16\left(\cos^2a_1\cos^2a_2\cos^2a_3 + \sin^2a_1\sin^2a_2\sin^2a_3\right).
\end{equation}
We can see that $0\leq e_p(U) \leq 2/9$. The CNOT gate with the nonlocal parameter $[\text{CNOT}] = [\pi/2,0,0]$ has the maximal entangling power $e_p(\text{CNOT}) = 2/9$. The SWAP gate with $[\text{SWAP}] = [\pi/2,\pi/2,\pi/2]$ has zero entangling power. Any two-qubit gate with nonzero entangling power can form a universal gate set with the single-qubit gates \cite{brylinski2002universal,bremner2002practical}.

Naturally, based on the value of the nonlocal parameters $[a_1,a_2,a_3]$, we can map each two-qubit gate on a point in the three-dimensional space, namely the 3-Torus (because of the periodicity of $[a_1,a_2,a_3]$). Even if we only consider the cube with length $\pi$, different points inside the cube may correspond to the two-qubit gates which are locally equivalent. 
For example, the two-qubit gate with $[\theta,0,0]$ is locally equivalent to the two-qubit gate with $[0,0,\theta]$ by applying the Hadamard transformation. 

\begin{figure}[t]
\centerline{\includegraphics[width=0.6\textwidth]{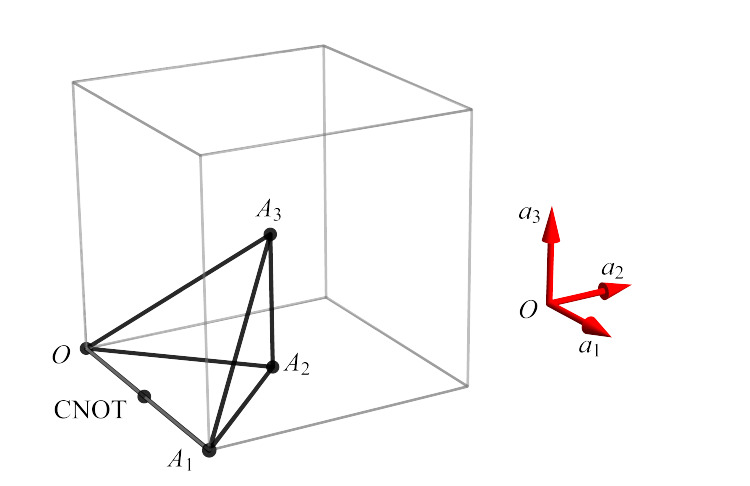}}
\caption{Geometric representation of the two-qubit gate $U=[a_1,a_2,a_3]$. The tetrahedron $OA_1A_2A_3$ has the vertexes $O=[0,0,0]$, $A_1=[\pi,0,0]$, $A_2=[\pi/2,\pi/2,0]$, and $A_3=[\pi/2,\pi/2,\pi/2]$.}
\label{fig:Geometric_two_qubit}
\end{figure}

One can apply the Weyl group to remove such ambiguity, called the Weyl chamber \cite{zhang2003geometric}. Then the cube can be divided into 24 Weyl chambers. Each chamber is a tetrahedron. Choose the tetrahedron with the vertexes $O=[0,0,0]$, $A_1=[\pi,0,0]$, $A_2=[\pi/2,\pi/2,0]$, and $A_3=[\pi/2,\pi/2,\pi/2]$, namely the region with $\pi-a_2\geq a_1\geq a_2\geq a_3\geq 0$. See Fig. \ref{fig:Geometric_two_qubit}. Different points in one Weyl chamber correspond to the different nonlocal two-qubit gates (except the points on the $OA_1A_2$ base). The middle of $OA_1$ represents the CNOT gate with $[\text{CNOT}] = [\pi/2,0,0]$. The SWAP gate is at the point $A_3$ with $[\text{SWAP}] = [\pi/2,\pi/2,\pi/2]$. Except for the vertexes $O$, $A_1$, and $A_3$, all two-qubit gates have nonzero entangling powers. Note that the two-qubit gates that have the same entangling powers may be not locally equivalent. 

\subsection{\label{subsec:decompose_two_qubit}Decomposition of two-qubit gate}

Quantum computers are equipped with several native gates. Then arbitrary unitary operation can be constructed or approximately constructed from those gates \cite{nielsenQuantumComputationQuantum2010}. How to construct arbitrary two-qubit gate is well studied \cite{zhangOptimalQuantumCircuit2004,vidalUniversalQuantumCircuit2004,vatanOptimalQuantumCircuits2004,shendeMinimalUniversalTwoqubit2004,zhang2005conditions}. The Cartan decomposition of two-qubit gate in Eq. (\ref{eq:U_decomposition}) naturally gives the following decomposition criterion \cite{zhangOptimalQuantumCircuit2004}.
\begin{theorem}
\label{theorem_minimal_decompose}
    Any two-qubit gate $[U] = [a_1,a_2,a_3]$ can be constructed from minimal $n$ ($n\geq 3$) applications of $U'(\theta)$ with $[U'(\theta)] = [\theta,0,0]$, if the nonlocal parameters satisfy $0\leq a_1+a_2+a_3\leq n\theta$ or $a_1-a_2-a_3\geq \pi-n\theta$.
\end{theorem}
Therefore any two-qubit gate can be constructed from at most three CNOTs. Two-qubit gates belong to $SO(4)$ have the nonlocal parameters $a_3 = 0$. Consequently, their decompositions only need at most two CNOTs \cite{vidalUniversalQuantumCircuit2004,vatanOptimalQuantumCircuits2004}. 

Applying the trick $\text{CNOT}(1\!\!1_2\otimes\sigma_z)\text{CNOT} = \sigma_z\otimes\sigma_z$, then we have
\begin{equation}
\label{eq:Rz_two_cnot}
    \Qcircuit @C=0.7em @R=1.5em {
    &&&&&&& \ctrl{1} & \qw & \ctrl{1} & \qw \\
    \raisebox{0.9cm}{$e^{-\frac i 2\theta(\sigma_z\otimes\sigma_z) }=$} &&&&&&& \targ & \gate{R_z(\theta)} & \targ & \qw 
    }
\end{equation}
with the single-qubit rotation gate $R_z(\theta)=e^{-\frac{i}{2}\theta\sigma_z}$. Applying the transformation $H\sigma_z H = \sigma_x$ or $\sqrt{\sigma_x}^\dag \sigma_z\sqrt{\sigma_x} = \sigma_y$ to $e^{-\frac i 2\theta(\sigma_z\otimes\sigma_z)}$ gives the realization of $e^{-\frac i 2\theta(\sigma_x\otimes\sigma_x)}$ or $e^{-\frac i 2\theta(\sigma_y\otimes\sigma_y)}$. Here $H$ is the Hadamard gate given by $H = (\sigma_x+\sigma_z)/\sqrt{2}$. Therefore, based on Eqs. (\ref{eq:U_decomposition}) and (\ref{eq:def_U_core}), we can decompose any two-qubit gate into six CNOTs. However, it is not optimal in the number of CNOTs according to Theorem \ref{theorem_minimal_decompose}. After some simplifications (by moving the single-qubit gates and merging the CNOTs), we have the optimal CNOT decomposition for any two-qubit gate (given by the universal gate set $\{R_z(\theta),H,\text{CNOT}\}$) \cite{zhangOptimalRealizationYang2024}
\begin{equation}
\label{eq:general_decompose_circuit}
    \Qcircuit @C=0.7em @R=1.5em {
    &&&& \gate{V_3} & \ctrl{1} & \gate{H} & \gate{S} & \gate{R^\dag_z\left(a_2\right)} & \ctrl{1} & \qw & \qw & \targ & \gate{S^\dag} & \gate{H} & \gate{V_1} & \qw \\
    \raisebox{1.1cm}{$U=$} &&&& \gate{V_4} & \targ & \gate{R^\dag_z(a_1)} & \qw & \qw & \targ & \gate{R_z(a_3)} & \gate{H} & \ctrl{-1} & \gate{S} & \gate{H} & \gate{V_2} &\qw 
    }
\end{equation}
with the phase gate $S = \text{diag}\{1,i\}$. Note that we have $S\cong R_z(\pi/2)$. The symbol $\cong$ represents the equivalent relation up to the global phase difference. The nonlocal parameters are interchangeable because of the commutativity of $\sigma_x\otimes\sigma_x$, $\sigma_y\otimes\sigma_y$, and $\sigma_z\otimes\sigma_z$. Note that the above decomposition is not optimal if $a_3=0$, since only two CNOTs are required.

\section{\label{sec:Geo_braid} Geometric representations and decompositions of braid gates}

\subsection{\label{subsec:braid_gate}Two-qubit braid gates}

The braid group relation or the constant Yang-Baxter equation characterizes the low-dimensional topology \cite{kauffman2001knots}. The topological entanglement can be described by the corresponding knot invariants. Then one can consider the unitary representation of the braid group. It works as a two-qubit gate, which can generate quantum entanglement. In the qubit representation, how to construct the braid matrix or the unitary braid gate is well studied \cite{hietarintaAllSolutionsConstant1992,dye2003unitary,Padmanabhan2020braidingquantum}.

The braid group has the generator $b_j$ with $j=1,2,\ldots,n-1$ satisfying 
\begin{subequations}
\begin{align}
    & b_jb_{j\pm 1}b_j =  b_{j\pm 1}b_j  b_{j\pm 1}, \\
    & b_jb_k = b_kb_j,\qquad |j-k|\geq 2.
\end{align}
\end{subequations}
In the context of quantum computation, we consider the braid group representation given by $b_j = 1\!\!1_2\otimes 1\!\!1_2\otimes \cdots \otimes B_{j,j+1} \otimes \cdots \otimes 1\!\!1_2\otimes 1\!\!1_2$, where $B_{j,j+1}$ is a two-qubit gate acting on the qubits $j$ and $j+1$. Here $B$ is called the braid gate. We omit the subscriptions in the following for simplicity. Note that the commutative relation of $b_j$ is satisfied automatically. The braid relation becomes 
\begin{equation}
\label{eq:braid_relation}
    (B\otimes 1\!\!1_2)(1\!\!1_2\otimes B)(B\otimes 1\!\!1_2) = (1\!\!1_2\otimes B)(B\otimes 1\!\!1_2)(1\!\!1_2\otimes B).
\end{equation}
Therefore, any two-qubit gate that satisfies the above relation is a braid gate. 

All solutions of $B$ in the two-qubit cases are known \cite{hietarintaAllSolutionsConstant1992,dye2003unitary}. Note that one has to add the unitary condition (to be a valid quantum gate). The core of the two-qubit gate $U_\text{core}$ defined in Eq. (\ref{eq:def_U_core}) is the X-type two-qubit gate. There are in total four X-type braid gates \cite{padmanabhan2021local}. In the two-qubit computational basis $\{|00\rangle,|01\rangle,|10\rangle,|11\rangle\}$, the four braid gates have the forms
\begin{subequations}
    \begin{equation}
    \label{eq:def_B_I}
        B_{I} = \begin{pmatrix}
        e^{i\varphi_1} & 0 & 0 & 0 \\
        0 & 0 & e^{i\varphi_2} & 0 \\
        0 & e^{i\varphi_3} & 0 & 0 \\
        0 & 0 & 0 & e^{i\varphi_4} \\
        \end{pmatrix},
    \end{equation}
    \begin{equation}
    \label{eq:def_B_II}
        B_{II} = \begin{pmatrix}
        0 & 0 & 0 & e^{i\varphi_2} \\
        0 & e^{i\varphi_1} & 0 & 0 \\
        0 & 0 & e^{i\varphi_1} & 0 \\
        e^{i\varphi_3} & 0 & 0 & 0 \\
        \end{pmatrix},
    \end{equation}
    \begin{equation}
    \label{eq:def_B_III}
        B_{III} = \begin{pmatrix}
        \cos\varphi_1 & 0 & 0 & \sin\varphi_1 e^{i\varphi_2} \\
        0 & -i\sin\varphi_1 & -\cos\varphi_1 & 0 \\
        0 & -\cos\varphi_1 & -i\sin\varphi_1 & 0 \\
        -\sin\varphi_1 e^{-i\varphi_2} & 0 & 0 & \cos\varphi_1 \\
        \end{pmatrix},
    \end{equation}
    \begin{equation}
    \label{eq:def_B_IV}
        B_{IV} = \frac{1}{\sqrt 2}\begin{pmatrix}
        1 & 0 & 0 & e^{i\varphi_1} \\
        0 & 1& 1 & 0 \\
        0 & -1 & 1 & 0 \\
        -e^{-i\varphi_1} & 0 & 0 & 1 \\
        \end{pmatrix}.
    \end{equation}
\end{subequations}
The parameters can be taken arbitrarily $\varphi_j\in[0,2\pi)$. In the following, we always take $\varphi_j$ as the braid gate parameters.

\subsection{\label{subsec:braid_gate_geo}Geometric representations and decompositions}

\begin{figure*}[t]
\centerline{\includegraphics[width=\textwidth]{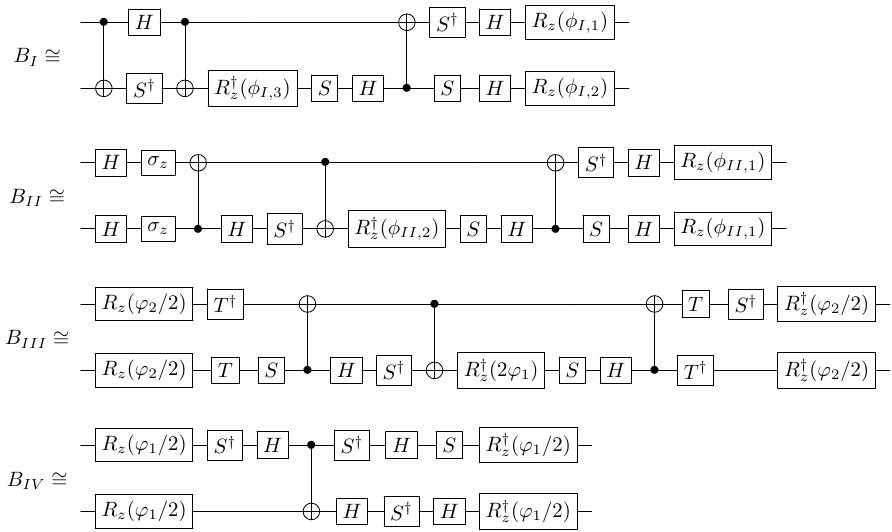}}
\caption{Decomposition of braid gates $B_I$ (\ref{eq:def_B_I}), $B_{II}$ (\ref{eq:def_B_II}), $B_{III}$ (\ref{eq:def_B_III}), and $B_{IV}$ (\ref{eq:def_B_IV}) into the universal gate set $\{R_z(\theta),H,\text{CNOT}\}$. The symbol $\cong$ means a global phase difference. The angles $\phi_I$ are defined in Eq. (\ref{eq:phi_I_1})-(\ref{eq:phi_I_3}). The angles $\phi_{II}$ are defined in Eq. (\ref{eq:phi_II_1})-(\ref{eq:phi_II_2}). Here $S$ is the phase gate with $S\cong R_z(\pi/2)$, and $T$ is the $\pi/8$ gate with $T\cong R_z(\pi/4)$.}
\label{fig:braid_gate_decomposition}
\end{figure*}

\begin{figure}[t]
\centerline{\includegraphics[width=0.5\textwidth]{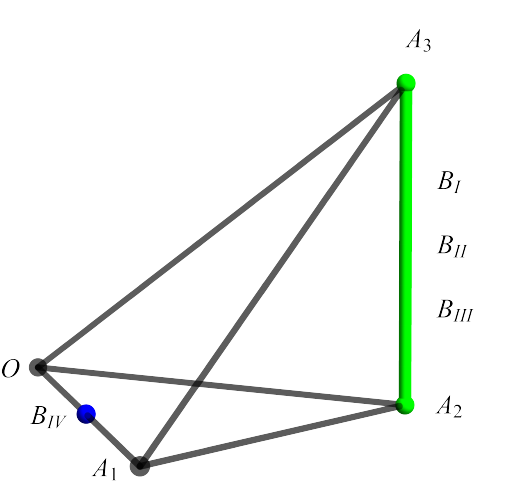}}
\caption{Geometric representation of braid gates $B_I$ (\ref{eq:def_B_I}), $B_{II}$ (\ref{eq:def_B_II}), $B_{III}$ (\ref{eq:def_B_III}), and $B_{IV}$ (\ref{eq:def_B_IV}).}
\label{fig:Geometric_braid_gate}
\end{figure}

Consider the universal gate set $\{R_z(\theta),H,\text{CNOT}\}$. Based on the standard decomposition of the two-qubit gate given in Eq. (\ref{eq:U_decomposition}), we have the decompositions of braid gates shown in Fig. \ref{fig:braid_gate_decomposition}. The parameters $\phi_I$ of braid gate $B_I$ are defined as
\begin{subequations}
    \begin{align}
    \label{eq:phi_I_1}
        \phi_{I,1} = &\frac 1 2\left(-\varphi_1-\varphi_2+\varphi_3+\varphi_4\right),\\
    \label{eq:phi_I_2}
        \phi_{I,2} = &\frac 1 2\left(-\varphi_1+\varphi_2-\varphi_3+\varphi_4\right),\\
    \label{eq:phi_I_3}
        \phi_{I,3} = &\frac 1 2\left(-\varphi_1+\varphi_2+\varphi_3-\varphi_4\right).
    \end{align}
\end{subequations}
The braid gate $B_{II}$ has the defined parameters
\begin{subequations}
    \begin{align}
    \label{eq:phi_II_1}
        \phi_{II,1} = & \frac 1 2\left(-\varphi_2+\varphi_3\right),\\
    \label{eq:phi_II_2}
        \phi_{II,2} = & \frac 1 2\left(-\varphi_2+2\varphi_1-\varphi_3\right).
    \end{align}
\end{subequations}
Note that both $B_{I}\text{SWAP}$ and $B_{II}\text{SWAP}(\sigma_x\otimes\sigma_x)$ are diagonal matrices.

\begin{table*}[t]
\small
    \centering
    \begin{tabular}{c|c|c|c} 
     \hline\hline
    & Clifford gate & Matchgate & ~Dual unitary gate~ \\ \hline
    $~B_{I}~$ & $\phi_{I,1},\phi_{I,2},\phi_{I,3}\in\{k\pi/2,k\in \mathbb Z\}$ & $\phi_{I,3} = \pi/2$ & Always \\
    $~B_{II}~$ & $\phi_{II,1},\phi_{II,2}\in\{k\pi/2,k\in \mathbb Z\}$ & $~\phi_{II,2} = \pi/2~$ & Always \\
    $~B_{III}~$ & $~\varphi_1\in \{k\pi/4,k\in \mathbb Z\}$ and $\varphi_2\in \{k\pi+\pi/2,k\in \mathbb Z\}~$ & Cannot & Always \\
    $~B_{IV}~$ & $\varphi_1\in \{k\pi,k\in \mathbb Z\}$ & Always & Cannot \\
\hline\hline
\end{tabular}
    \caption{Conditions for braid gates $B_I$ (\ref{eq:def_B_I}), $B_{II}$ (\ref{eq:def_B_II}), $B_{III}$ (\ref{eq:def_B_III}), and $B_{IV}$ (\ref{eq:def_B_IV}) to be the Clifford, the matchgate, and the dual unitary gate.}
    \label{tab:braid_conditions}
\end{table*}

The braid gates have the nonlocal parameters
\begin{subequations}
\begin{align}
     [B_I] = & \frac 1 2 \left[\pi,\pi,\pi- 2\phi_{I,3}\right], \\
     [B_{II}] = & \frac 1 2\left[\pi,\pi,\pi-2\phi_{II,2}\right], \\
     [B_{III}] = &\frac 1 2\left[\pi,\pi,\pi-4\varphi_1\right],\\
     [B_{IV}] = &\frac 1 2 \left[\pi,0,0\right].
\end{align}
\end{subequations}
In the viewpoint of integrable vertex models, $B_I$ and $B_{II}$ can be categorized as the six-vertex model, while $B_{III}$ and $B_{IV}$ are from the eight-vertex model. However, in the context of quantum computation, the braid gates $B_{I,II,III}$ are all on the edge $A_2A_3$ of the two-qubit tetrahedron. See Fig. \ref{fig:Geometric_braid_gate}. Correspondingly, their gate decompositions are similar and require minimal three CNOTs if $B_{I,II,III}$ are not at the point $A_2$. See Fig. \ref{fig:braid_gate_decomposition}. The braid gate $B_{IV}$ is a special case that is equivalent to the CNOT.

It is straightforward to calculate the entangling powers of the four braid gates. Specifically, we have
\begin{subequations}
    \begin{align}
        e_p(B_I) = & \frac 2 9 \sin^2\phi_{I,3}, \\
        e_p(B_{II}) = & \frac 2 9 \sin^2\phi_{II,2}, \\
        e_p(B_{III}) = & \frac 2 9 \sin^2\left(2\varphi_1\right),\\
        e_p(B_{IV}) = & \frac 2 9.
    \end{align}
\end{subequations}
Braid gates $B_{I,II,III}$ have the maximal entangling powers when $\phi_{I,3} = \pi/2$, or $\phi_{II,2} =\pi/2$, or $\varphi_1 =\pi/4$, which are locally equivalent to the iSWAP gate defined in Eq. (\ref{eq:def_iswap}). When braid gates $B_{I,II,III}$ (with $\phi_{I,3} = 0$, or $\phi_{II,2} =0$, or $\varphi_1 =0$) are locally equivalent to the SWAP gate, they have zero entangling power. Here we can see the difficulties of characterizing the quantum entanglement via the topological invariants because not all two-qubit gates are locally equivalent to a braid gate.

Clifford gates are elements of the Clifford group, which preserves the Pauli group, therefore circuits only composed of Clifford gates can be classically simulated \cite{gottesmanStabilizerCodesQuantum1997,gottesmanTheoryFaulttolerantQuantum1998}. Another equivalent definition of the Clifford gate is that the quantum gate can be decomposed with $\{H,S,\text{CNOT}\}$. Based on the gate decomposition of the four braid gates, we know that the braid gate $B_I$ is a Clifford gate if $\phi_{I,1},\phi_{I,2},\phi_{I,3}\in\{k\pi/2,k\in \mathbb Z\}$; the braid gate $B_{II}$ is a Clifford gate if $\phi_{II,1},\phi_{II,2}\in\{k\pi/2,k\in \mathbb Z\}$; the braid gate $B_{III}$ is a Clifford gate if $\varphi_1\in \{k\pi/4,k\in \mathbb Z\}$ and $\varphi_2\in \{k\pi+\pi/2,k\in \mathbb Z\}$; the braid gate $B_{IV}$ is a Clifford gate if $\varphi_1\in \{k\pi,k\in \mathbb Z\}$. We summarize the results in Table \ref{tab:braid_conditions}.

Matchgate is a special family of X-type two-qubit gates. Quantum circuits only composed of the matchgates acting on the nearest neighbor qubits is another family of circuits that is classically tractable \cite{valiantQuantumComputersThat2001,jozsaMatchgatesClassicalSimulation2008}. Matchgate circuit can be mapped to the dynamics of the free fermion model \cite{terhalClassicalSimulationNoninteractingfermion2002}. The necessary condition of being a matchgate is to have a vanishing nonlocal parameter \cite{brodExtendingMatchgatesUniversal2011}. We can check that the braid gate $B_I$ is a matchgate if $\phi_{I,3} = \pi/2$; the braid gate $B_{II}$ is a matchgate if $\phi_{II,2} = \pi/2$; the braid gate $B_{III}$ can not be a matchgate; the braid gate $B_{IV}$ is always a matchgate.

It is interesting to see that two-qubit gates on the edge $A_2A_3$ belong to the so-called dual-unitary gate \cite{bertiniExactCorrelationFunctions2019}. All quantum gates are unitary. The dual unitary gate satisfies an additional unitary condition $\tilde U \tilde U^\dag = \tilde U^\dag \tilde U = 1\!\!1$, where $\tilde U$ comes from the index change $\langle m|\otimes \langle n|\tilde U|i\rangle\otimes|j\rangle = \langle j|\otimes \langle n|U|i\rangle\otimes |m\rangle$ where $|i\rangle$, $|j\rangle$, $|m\rangle$, and $|n\rangle$ are all the computational basis. The dynamical correlation functions of the brick-wall circuit composed of the dual-unitary gate can be exactly calculated. From Fig. \ref{fig:Geometric_braid_gate}, we see that braid gates $B_I$, $B_{II}$, and $B_{IV}$ are all dual-unitary, while $B_{IV}$ is not. 


\section{\label{sec:YBG} Geometric representations and decompositions of Yang-Baxter gates}

\subsection{\label{subsec:Baxterization}Yang-Baxterization}

In the braid relation (\ref{eq:braid_relation}), the braid parameters are all the same. The braid relation can be generalized to a parameter-dependent relation, given by\footnote{The $R$ matrix acting on the adjacent qubits is commonly denoted as $\check R$. Here we omit the check mark for simplicity.}
\begin{equation}
\label{eq:YBE}
    (R(x)\otimes 1\!\!1_2)(1\!\!1_2\otimes R(xy))(R(y)\otimes 1\!\!1_2) = (1\!\!1_2\otimes R(y))(R(xy)\otimes 1\!\!1_2)(1\!\!1_2\otimes R(x)).
\end{equation}
It is called the Yang-Baxter equation, where $x$ is called the spectral parameter \cite{jimbo1990yang,Perk2006}. The Yang-Baxter equation is the key in the quantum inverse scattering method which systematically solves the Yang-Baxter integrable models \cite{korepin1997quantum}. 
    
Different $R$ matrices correspond to different integrable models. Finding solutions of the Yang-Baxter equation is an important subject in mathematical physics. There are various ways to construct the $R$ matrices. One method is starting from the braid matrix, and then introducing the spectral parameters in the construction. Such procedure is called the Yang-Baxterization \cite{jonesBaxterization1990,chengYangBaxterizationBraidGroup1991,geExplicitTrigonometricYangbaxterization1991}. The idea of Yang-Baxterization is to replace the eigenvalues of the braid matrix with the polynomial of the spectral parameters. See \cite{zhang2005universal,zhang2005yang} for the Yang-Baxterization examples. 

Suppose that the braid matrix has the spectral decompositions
\begin{equation}
    B = \sum_{j=1}^n \lambda_jP_j,
\end{equation}
with $n$ distinct eigenvalues $\lambda_j$ and the corresponding projectors $P_j$. The ansatz of $R$ matrix from the Yang-Baxterization is
\begin{equation}
    R(x) = \sum_{j=1}^n \Theta^{(n)}_j(x)P_j,
\end{equation}
with the polynomial of $x$ in the degree of $n-1$. If $n=2$, we have
\begin{subequations}
    \begin{align}
    \Theta^{(2)}_1(x) = & x + \frac{\lambda_1}{\lambda_2}, \\
    \Theta^{(2)}_2(x) = & 1 + x\frac{\lambda_1}{\lambda_2},
\end{align}
\end{subequations}
which gives
\begin{equation}
    R(x) = \frac{1}{\lambda_2}\left(B+x\lambda_1\lambda_2B^{-1}\right).
\end{equation}
The exchange on $\lambda_1\leftrightarrow\lambda_2$ gives the same $R$ matrix. If $n=3$, we have
\begin{subequations}
    \begin{align}
        \Theta^{(3)}_1(x) = & \left(x+\frac{\lambda_1}{\lambda_2}\right)\left(x+\frac{\lambda_2}{\lambda_3}\right), \\
        \Theta^{(3)}_2(x) = & \left(1+x\frac{\lambda_1}{\lambda_2}\right)\left(x+\frac{\lambda_2}{\lambda_3}\right), \\
        \Theta^{(3)}_3(x) = & \left(1+x\frac{\lambda_1}{\lambda_2}\right)\left(1+x\frac{\lambda_2}{\lambda_3}\right).
    \end{align}
\end{subequations}
The corresponding $R(x)$ matrix can be rewritten as
\begin{equation}
\label{eq:R_x_three_eigenvalues}
    R(x) = \alpha(x,\lambda_3)B + \beta(x,\lambda_1,\lambda_2,\lambda_3)1\!\!1 + \gamma(x,\lambda_1) B^{-1},
\end{equation}
with
\begin{subequations}
\begin{align}
    & \alpha(x,\lambda_3) =  -\frac{1}{\lambda_3}(x-1),\\
    & \beta(x,\lambda_1,\lambda_2,\lambda_3) =  \left(1+\frac{\lambda_1}{\lambda_2} + \frac{\lambda_1}{\lambda_3} + \frac{\lambda_2}{\lambda_3}\right)x, \\
    & \gamma(x,\lambda_1) =  \lambda_1x(x-1).
\end{align}    
\end{subequations}
The exchange on $\lambda_1\leftrightarrow\lambda_2$ or $\lambda_2\leftrightarrow\lambda_3$ may give different $R$ matrices. We emphasize that the Yang-Baxterization only gives ansatz of the $R$ matrix if $n\geq 3$. Additional constraints on the braid matrix are required. For example, if the braid matrix with three distinct eigenvalues admits the Birman–Murakami–Wenzl (BMW) algebra \cite{birmanBraidsLinkPolynomials1989,murakamiKauffmanPolynomialLinks1987}, the Yang-Baxterization always works \cite{chengYangBaxterizationBraidGroup1991,geExplicitTrigonometricYangbaxterization1991}. 

Since we are only interested in the $R$ matrix which can act as a quantum gate, the $R$ matrix has to be unitary. For the braid gate with two distinct eigenvalues, we have 
\begin{theorem}
\label{theorem_real_x}
If the braid gate $B$ has two distinct eigenvalues and satisfying $B^2\neq 1\!\!1$, the $R$ matrix obtained from the Yang-Baxterization on $B$ is unitary if the spectral parameter is real.
\end{theorem}
The proof is straightforward and can be found in \ref{App:A}. If $B^2=1\!\!1$, such as $B=\text{SWAP}$, the Yang-Baxterization gives a trivial $R(x)$ matrix where the spectral parameter $x$ is only shown in the global phase.

\subsection{\label{subsec:YBG}Yang-Baxter gates from Yang-Baxterization}

The braid gate $B_I$ has the eigenvalues
\begin{subequations}
    \begin{align}
        \lambda_{I,1} = & -e^{\frac i 2(\varphi_2+\varphi_3)},\\
        \lambda_{I,2} = & e^{i\varphi_1}, \\
        \lambda_{I,3} = & e^{\frac i 2(\varphi_2+\varphi_3)}, \\
        \lambda_{I,4} = & e^{i\varphi_4}.
    \end{align}
\end{subequations}
If we consider the Yang-Baxterization with four distinct eigenvalues, it does not give the correct $R$ matrix. Therefore we restrict to $\varphi_1 = \varphi_4$, namely three distinct eigenvalues. The braid gate $B_I$ with $\varphi_1 = \varphi_4$ has a simple relation with $B_{II}$ given by
\begin{equation}
\label{eq:B_I_and_B_II}
    B_I(\varphi_1 = \varphi_4) = (1\!\!1_2\otimes\sigma_x)B_{II}(1\!\!1_2\otimes\sigma_x).
\end{equation}
It follows that they have the same eigenvalues, denoted as
\begin{subequations}
    \begin{align}
        \lambda_{II,1} = & -e^{\frac i 2(\varphi_2+\varphi_3)},\\
        \lambda_{II,2} = & e^{i\varphi_1}, \\
        \lambda_{II,3} = & e^{\frac i 2(\varphi_2+\varphi_3)}.
    \end{align}
\end{subequations}
Note that the $R(x)$ matrices obtained from Eq. (\ref{eq:R_x_three_eigenvalues}) preserves the transformation, since Pauli matrix $\sigma_x$ is Hermitian. 

Consider the braid gate $B_{III}$, which has the eigenvalues
\begin{subequations}
    \begin{align}
        \lambda_{III,1} = & -e^{i\varphi_1}, \\
        \lambda_{III,2} = & e^{-i\varphi_1}, \\
        \lambda_{III,3} = & e^{i\varphi_1}.
    \end{align}
\end{subequations}
The eigenvalues are only dependent on the braid parameter $\varphi_1$.

The Yang-Baxterization of three distinct eigenvalues, namely Eq. (\ref{eq:R_x_three_eigenvalues}) gives the following $R$ matrices
\begin{equation}
\label{eq:R_1}
    R_{l,1}(x) = \alpha(x,\lambda_{l,3})B_l \\
    + \beta(x,\lambda_{l,1},\lambda_{l,2},\lambda_{l,3})1\!\!1 + \gamma(x,\lambda_{l,1}) B^\dag_l,
\end{equation}
with $l\in\{I,II,III\}$. We call them the first kind of $R$ matrix (or Yang-Baxter gate) of $B_l$. For $B_{III}$, it admits a representation of BMW algebra, therefore the solution of the Yang-Baxter equation is guaranteed. Although the braid gates $B_{I}$ and $B_{II}$ do not have the BMW algebra representation, they satisfy the necessary condition for Yang-Baxterization given in \cite{geExplicitTrigonometricYangbaxterization1991}, therefore $R_{II,1}$ and $R_{II,2}$ also satisfy the Yang-Baxter equation.  

We observe that the eigenvalues of $B_{I}$, $B_{II}$, and $B_{III}$ both satisfy
\begin{subequations}
    \begin{align}   \beta(x,\lambda_{l,2},\lambda_{l,1},\lambda_{l,3}) = 0,\\
    \beta(x,\lambda_{l,1},\lambda_{l,3},\lambda_{l,2}) = 0,
    \end{align}
\end{subequations}
with $l\in\{I,II,III\}$. Here $\beta$ is the coefficient of the Yang-Baxterization in Eq. (\ref{eq:R_x_three_eigenvalues}). Therefore, the three-eigenvalue Yang-Baxterization reduces to the two-eigenvalue Yang-Baxterization. Specifically, we have another two types of $R$ matrices
\begin{subequations}
    \begin{align}
        \label{eq:R_2}
        R_{l,2}(x) = & \alpha(x,\lambda_{l,3})B_l + \gamma(x,\lambda_{l,2}) B^\dag_l, \\
        \label{eq:R_3}
        R_{l,3}(x) = & \alpha(x,\lambda_{l,2})B_l + \gamma(x,\lambda_{l,1}) B^\dag_l,
    \end{align}
\end{subequations}
with $l\in\{I,II,III\}$. We call them the second and third kinds of $R$ matrices (or Yang-Baxter gate) of $B_l$. Note that the second and third kinds of Yang-Baxter gates are obtained from the eigenvalue permutation $\lambda_1\leftrightarrow\lambda_2$ and $\lambda_2\leftrightarrow\lambda_3$ of the first kind. The unitary condition requires that the spectral parameter in the first kind $R_{l,1}$ is real. From Theorem \ref{theorem_real_x}, we know that $R_{l,2}$ and $R_{l,3}$ are unitary if the spectral parameter $x$ is also real, therefore work as two-qubit gates. 

Specifically, $R_{I,1}$ is the $R$ matrix of the XXZ model \cite{doikou2010introduction}. The braid gate parameter in $R_{I,1}$ corresponds to the anisotropic coupling of the spin chain. The $R_{III,1}$ matrix can also be obtained from the trigonometric limits of elliptic $R$ matrix \cite{arnaudon2000towards}. Moreover, the $R_{III,1}$ matrix is locally equivalent to the $R$ matrix of the sine-Gordon model \cite{wen2001modular}. The matrix expressions of $R_{l,1}$ $R_{l,2}$, and $R_{l,3}$, namely total nine $R$ gates, can be found in \ref{App:B}. 

The braid gate $B_{IV}$ only has two distinct eigenvalues, $\lambda_{IV,1} = e^{i\pi/4}$ and $\lambda_{IV,2} = e^{-i\pi/4}$, therefore the Yang-Baxterization gives a unique $R$ matrix, namely
\begin{equation}
\label{eq:R_IV}
    R_{IV}(\chi) \cong \begin{pmatrix}
        \cos\chi & 0 & 0 & e^{i\varphi_1}\sin\chi \\
        0 & \cos\chi & \sin\chi & 0 \\
        0 & -\sin\chi & \cos\chi & 0 \\
        -e^{-i\varphi_1}\sin\chi & 0 & 0 & \cos\chi
        \end{pmatrix},
\end{equation}
with the real spectral parameter \footnote{It is a non-conventional way to parameterize the spectral parameter. Therefore we denote it as $\chi$, to distinguish it from $\mu$.}
\begin{equation}
\label{eq:spectral_chi}
    x = \tan(\chi - \pi/4).
\end{equation}
The optimal gate and pulse constructions of $R_{IV}(\chi)$ with $\varphi_1=0$ have been studied in detail in  \cite{zhangOptimalRealizationYang2024}.

\subsection{\label{subsec:YBG_geo}Geometric representations of Yang-Baxter gates}

\begin{figure*}[t]
\centerline{\includegraphics[width=\textwidth]{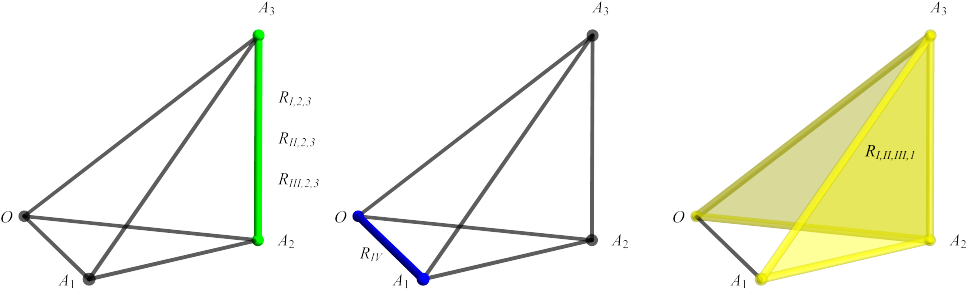}}
\caption{Geometric representations of Yang-Baxter gates obtained from braid gates. Yang-Baxter gates $R_{l,1}$, $R_{l,2}$, and $R_{l,3}$ with $l\in\{I,II,III\}$ are defined in Eqs. (\ref{eq:R_1}), (\ref{eq:R_2}), and (\ref{eq:R_3}) respectively. The Yang-Baxter gate $R_{IV}$ is given in Eq. (\ref{eq:R_IV}).}
\label{fig:geo_YBG}
\end{figure*}

Through some analysis, we find the nonlocal parameters of $R_{I,1}$ are
\begin{equation}
\label{eq:R_I_1_nonlocal}
    [R_{I,1}] = [a_{I,1},a_{I,1},c_{I,1}],
\end{equation}
with
\begin{subequations}
    \begin{align}
    \label{eq:a_I_1}
        a_{I,1} = & \arccos\left(\sqrt{\frac{1-\cos 2\varphi}{\cosh 2\mu - \cos2\varphi}}\right), \\
    \label{eq:c_I_1}
        c_{I,1} = & \frac{i}{2}\ln \left(\frac{\sin(\varphi+i\mu)}{\sin(\varphi-i\mu)}\right).
    \end{align}
\end{subequations}
Here $\mu = \ln x$ with $x>0$ and $\varphi = (\varphi_2+\varphi_3)/2-\varphi_1$. Therefore, the nonlocal parameters are jointly determined by the spectral parameter $\mu$ and the braid gate parameter $\varphi$. The Yang-Baxter gates obtained from $B_I$ are locally equivalent to the ones obtained from $B_{II}$, suggested by Eq. (\ref{eq:B_I_and_B_II}). Therefore, we have 
\begin{equation}
    [R_{II,1}] = [R_{I,1}].
\end{equation}
Note that $a_{I,1}\in[0,\pi/2]$ and $c_{I,1}\in[-\pi/2,\pi/2]$. When $\mu\rightarrow -\infty$, we restore the braid gate $B_I$, corresponding to the boundary condition $R(x\rightarrow0) \rightarrow B$. 

The two-qubit tetrahedron requires the range $\pi-a_{I,1}\geq a_{I,1}\geq c_{I,1}\geq 0$. We can always gauge the nonlocal parameters in the two-qubit tetrahedron through the local transformations. For example, if $c_{I,1}<0$, we have the two-qubit tetrahedron $[\pi-a_{I,1},a_{I,1},-c_{I,1}]$. See Fig. \ref{fig:geo_YBG}. We can see that $R_{I,1}$ and $R_{I,2}$ are on the faces of $OA_1A_2$ and $OA_2A_3$.

The Yang-Baxter gates $R_{I,2}$ and $R_{I,3}$ are SWAP-like gates (exchanging states $|01\rangle$ and $|10\rangle$ with some phases). See the \ref{App:B}. Therefore, they have similar geometric representations as $B_I$. Specifically, they have the nonlocal parameters
\begin{subequations}
    \begin{align}
        [R_{I,2}] = & \left[\frac{\pi}{2},\frac{\pi}{2},\frac{\pi}{2}-\varphi_{I,2}\right], \\
        [R_{I,3}] = & \left[\frac{\pi}{2},\frac{\pi}{2},\frac{\pi}{2}-\varphi_{I,3}\right],
    \end{align}
\end{subequations}
with
\begin{subequations}
    \begin{align}
    \label{eq:c_I_2}
        \varphi_{I,2} = & \frac 1 i \ln\left(\frac{\sin\left(\frac 1 2(\varphi-i\mu)\right)}{\sin\left(\frac 1 2(\varphi+i\mu)\right)}\right), \\
    \label{eq:c_I_3}
        \varphi_{I,3} = & \frac 1 i \ln\left(\frac{\cos\left(\frac 1 2(\varphi+i\mu)\right)}{\cos\left(\frac 1 2(\varphi-i\mu)\right)}\right).
    \end{align}
\end{subequations}
Since the two nonlocal parameters of $R_{I,2}$ and $R_{I,3}$ are constant, we can always gauge the third in the range $[0,\pi/2]$. Therefore, $R_{I,2}$ and $R_{I,3}$ are on the edge $A_2A_3$ of the two-qubit tetrahedron. Moreover, because of Eq. (\ref{eq:B_I_and_B_II}), we have 
\begin{subequations}
    \begin{align}
        [R_{II,2}] = [R_{I,2}], \\
        [R_{II,3}] = [R_{I,3}].
    \end{align}
\end{subequations}
See Fig. \ref{fig:geo_YBG} for the two-qubit tetrahedron of $R_I$ and $R_{II}$.

It is expected that the Yang-Baxter gate $R_{III}$ is locally equivalent to $R_{I}$ and $R_{II}$, since $B_I$, $B_{II}$, and $B_{III}$ have similar geometric representations (all on the edge $A_2A_3$ of the two-qubit tetrahedron). Through some calculations, we find
\begin{equation}
    [R_{III,1}] = [a_{III,1},a_{III,1},c_{III,1}],
\end{equation}
with
\begin{subequations}
    \begin{align}
    \label{eq:a_III_1}
        a_{III,1} = & \arccos\left(\sqrt{\frac{1-\cos 4\varphi_1}{\cosh 4\mu - \cos4\varphi_1}}\right), \\
    \label{eq:c_III_1}
        c_{III,1} = & \frac{i}{2}\ln \left(\frac{\sin(2\varphi_1+2i\mu)}{\sin(2\varphi_1-2i\mu)}\right).
    \end{align}
\end{subequations}
A simple comparison with the nonlocal parameters of $R_{III,1}$ and $R_{I,1}$ in Eq. (\ref{eq:R_I_1_nonlocal}) suggests that $R_{I,1}$ and $R_{III,1}$ have the correspondence through $2\varphi_1 \leftrightarrow \varphi$ and $2\mu\leftrightarrow \mu$. Therefore, $R_{III,1}$ is also on the faces $OA_1A_2$ and $OA_2A_3$ of the two-qubit tetrahedron. See Fig. \ref{fig:geo_YBG}.

The second and third kinds of $R_{III}$ have the nonlocal parameters
\begin{subequations}
    \begin{align}
        [R_{III,2}] = & \left[\frac{\pi}{2},\frac{\pi}{2},\frac{\pi}{2} - 2\varphi_{III,2}\right], \\
        [R_{III,3}] = & \left[\frac{\pi}{2},\frac{\pi}{2},\frac{\pi}{2} - 2\varphi_{III,3}\right],
    \end{align}
\end{subequations}
with
\begin{subequations}
    \begin{align}
    \label{eq:varphi_III_2}
        \varphi_{III,2} = & \arccos\left(\frac{\sinh\mu\cos\varphi_1}{\sqrt{\sinh^2\mu\cos^2\varphi_1 + \cosh^2\mu\sin^2\varphi_1}}\right), \\
    \label{eq:varphi_III_3}
        \varphi_{III,3} = & \arccos\left(\frac{\cosh\mu\cos\varphi_1}{\sqrt{\cosh^2\mu\cos^2\varphi_1 + \sinh^2\mu\sin^2\varphi_1}}\right).
    \end{align}
\end{subequations}
Therefore, $R_{III,2}$ and $R_{III,3}$ are on the edge $A_2A_3$ of the two-qubit tetrahedron. See Fig. \ref{fig:geo_YBG}. When $\mu = 0$, $R_{III,2}$ becomes locally equivalent to the SWAP gate. While $R_{III,3}$ requires $\varphi_1 = \pi/2$ to be locally equivalent to the SWAP gate. 

The Yang-Baxter gate $R_{IV}$ generated from $B
_{IV}$ has the nonlocal parameters
\begin{equation}
    [R_{IV}] = [2\chi,0,0],
\end{equation}
with the spectral parameter $\chi$ defined in Eq. (\ref{eq:spectral_chi}). It represents the edge $OA_1$ of the two-qubit tetrahedron. See Fig. \ref{fig:geo_YBG}. From Eq. (\ref{eq:R_IV}), we can easily see that $R_{IV}(\chi = 0)\cong 1\!\!1_4$ and $R_{IV}(\chi = \pi/2)$ is locally equivalent to $\sigma_x\otimes\sigma_x$.

\subsection{\label{subsec:YBG_decomposition}Gate decompositions of Yang-Baxter gates}

\begin{figure*}[t]
\centerline{\includegraphics[width=\textwidth]{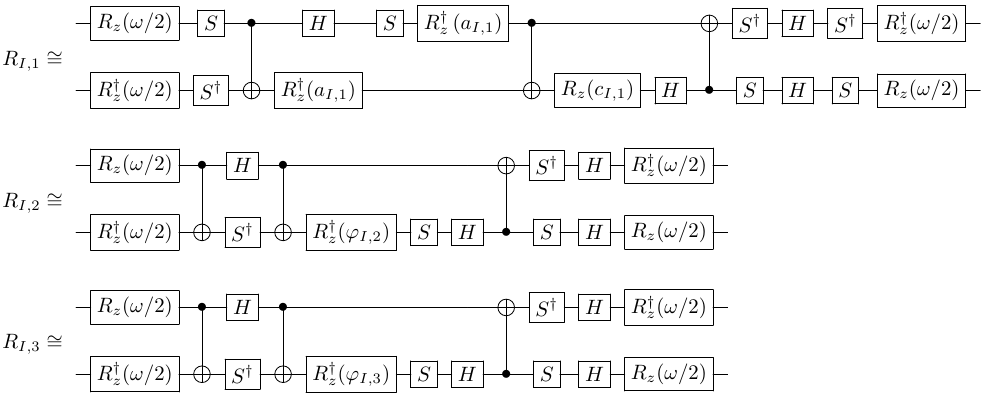}}
\caption{Decomposition of Yang-Baxter gates $R_{I,1}$ (\ref{eq:R_1}), $R_{I,2}$ (\ref{eq:R_2}), and $R_{I,3}$ (\ref{eq:R_3}) into the universal gate set $\{R_z(\theta),H,\text{CNOT}\}$. The parameters $a_{I,1}$, $c_{I,1}$, $\varphi_{I,2}$, and $\varphi_{I,3}$ are nonlocal parameters given in Eqs. (\ref{eq:a_I_1}), (\ref{eq:c_I_1}), (\ref{eq:c_I_2}), and (\ref{eq:c_I_3}) respectively. The parameter $\omega$ is given in Eq. (\ref{eq:def_R_I_omega}).}
\label{fig:R_I_decomposition}
\end{figure*}

\begin{figure*}[t]
\centerline{\includegraphics[width=\textwidth]{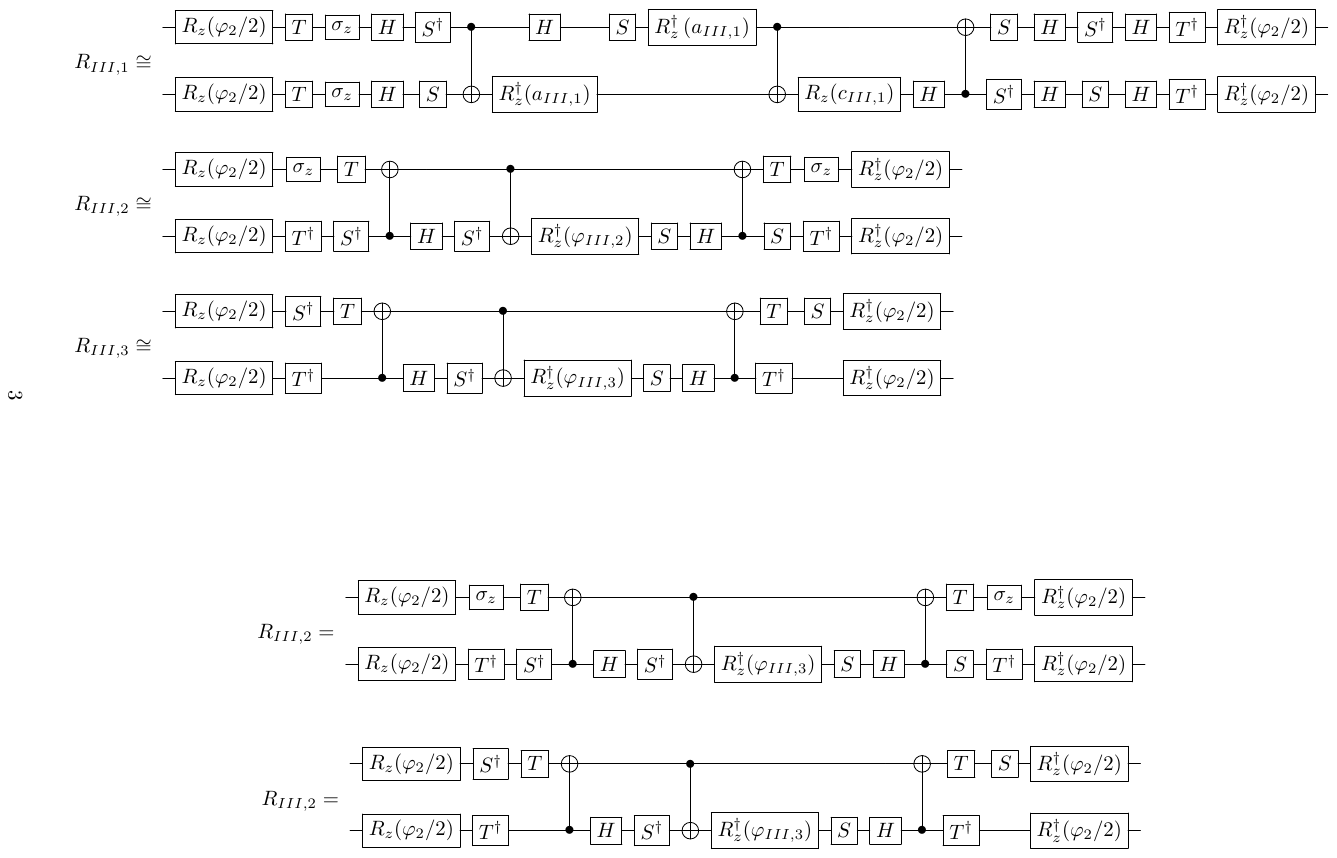}}
\caption{Decomposition of Yang-Baxter gates $R_{III,1}$ (\ref{eq:R_1}), $R_{III,2}$ (\ref{eq:R_2}), and $R_{III,3}$ (\ref{eq:R_3}) into the universal gate set $\{R_z(\theta),H,\text{CNOT}\}$. The parameters $a_{III,1}$, $c_{III,1}$, $\varphi_{III,2}$, and $\varphi_{III,2}$ are nonlocal parameters given in Eqs. (\ref{eq:a_III_1}), (\ref{eq:c_III_1}), (\ref{eq:varphi_III_2}), and (\ref{eq:varphi_III_3}) respectively.}
\label{fig:R_III_decomposition}
\end{figure*}

Recall that the general two-qubit gate can be decomposed into three CNOTs if its three nonlocal parameters are known. Based on the nonlocal parameters of the Yang-Baxter gates presented in the last subsection, we find the optimal gate decomposition (with the minimal number of CNOTs) of $R_I$ and $R_{III}$ in terms of the universal gate set $\{R_z(\theta),H,\text{CNOT}\}$. See Figs. \ref{fig:R_I_decomposition} and \ref{fig:R_III_decomposition}. For $R_I$, we define the parameter $\omega$ as
\begin{equation}
\label{eq:def_R_I_omega}
    \omega = \frac 1 2 (\varphi_2-\varphi_3),
\end{equation}
which only appears in the single-qubit gates of the decomposition. It is also called the $q$ deformation parameter of the $R$ matrix \cite{doikou2010introduction}. In practice, if the CNOT is not the native two-qubit gate of the quantum computer, the Yang-Baxter gate which is closest to such native two-qubit gate on the two-qubit tetrahedron has some advantages in realizations. For example, if the quantum computer has the native iSWAP gate, then one can construct the Yang-Baxter gate (with a specific spectral parameter) with only one iSWAP gate with some additional single-qubit gates.

Note that $R_I$ and $R_{II}$ have the relation
\begin{equation}
\label{eq:R_I_to_R_II}
    R_{I,1,2,3} = (1\!\!1_2\otimes\sigma_x)R_{II,1,2,3}(1\!\!1_2\otimes\sigma_x),
\end{equation}
suggested by Eq. (\ref{eq:B_I_and_B_II}), therefore the gate decompositions of $R_{II}$ are almost identical with $R_{I}$. When $R_{l,2,3}$ with $l\in\{I,II,III\}$ are at $A_2$ of the two-qubit tetrahedron, their decompositions only require two CNOTs. Similarly, when $R_{l,1}$ with $l\in\{I,II,III\}$ are at the edge $OA_2$ or $A_1A_2$ (with one vanishing nonlocal parameter), only two CNOTs are required.

Notice that the Yang-Baxter gate $R_{IV}$ has two vanishing nonlocal parameters, therefore at most two CNOTs are needed for its construction. The optimal decomposition of $R_{IV}$ is 
\begin{equation}
    \Qcircuit @C=0.4em @R=1.5em {
     &&&&&& \gate{R_z(\varphi_1/2)} & \gate{S} & \gate{H} & \ctrl{1} & \qw & \ctrl{1} & \gate{H} & \gate{S^\dag} & \gate{R^\dag_z(\varphi_1/2)} & \qw \\
     \raisebox{1.3cm}{$R_{IV}=$} &&&&&& \gate{R_z(\varphi_1/2)} & \gate{H} & \qw & \targ & \gate{R_z(2\chi)} & \targ & \gate{H} & \qw & \gate{R^\dag_z(\varphi_1/2)} & \qw
    }
\end{equation}
with the braid gate parameter $\varphi_1$ and the spectral parameter $\chi$. The gate identity (\ref{eq:Rz_two_cnot}) has been applied.  

\begin{table*}[t]
\small
    \centering
    \begin{tabular}{c|c|c|c} 
     \hline\hline
    & Clifford gate & Matchgate & ~Dual unitary gate~ \\ \hline
    \multirow{2}*{$R_{I,1}$} & $\omega\in\{k\pi,k\in\mathbb Z\}$ & \multirow{2}*{$\varphi = \pi/2$} & \multirow{2}*{$\mu\rightarrow -\infty$} \\
    & $\mu = 0$ or $\varphi\in \{k\pi,k\in\mathbb Z\}$ & & \\\hline
    \multirow{2}*{$R_{I,2}$} & $\omega\in\{k\pi,k\in\mathbb Z\}$ & \multirow{2}*{~Eq. (\ref{eq:Clifford_condition_R_I}) is satisfied~} & \multirow{2}*{Always} \\
    & $\mu = 0$ or $\varphi\in \{k\pi,k\in\mathbb Z\}$ or Eq. (\ref{eq:Clifford_condition_R_I}) is satisfied &  & \\\hline
    \multirow{2}*{$R_{I,3}$} & $\omega\in\{k\pi,k\in\mathbb Z\}$ & \multirow{2}*{Eq. (\ref{eq:Clifford_condition_R_II}) is satisfied} & \multirow{2}*{Always} \\
    & $\mu = 0$ or $\varphi\in \{k\pi,k\in\mathbb Z\}$ or Eq. (\ref{eq:Clifford_condition_R_II}) is satisfied & & \\\hline
    \multirow{2}*{$R_{III,1}$} & $\varphi_2\in\{k\pi+\pi/2,k\in\mathbb Z\}$ & \multirow{2}*{Cannot} & \multirow{2}*{$\mu\rightarrow -\infty$} \\
    & $\mu = 0$ or $\varphi_1\in \{k\pi/2,k\in\mathbb Z\}$ & & \\\hline
    \multirow{2}*{$R_{III,2}$} & $\varphi_2\in\{k\pi+\pi/2,k\in\mathbb Z\}$ & \multirow{2}*{Cannot} & \multirow{2}*{Always} \\
    & $\mu=0$ or $\varphi_1 \in\{k\pi+\pi/2,k\in\mathbb Z\}$ or $\varphi_1 \in\{k\pi,k\in\mathbb Z\}$ & & \\\hline
    ~\multirow{2}*{$R_{III,3}$}~ & $\varphi_2\in\{k\pi+\pi/2,k\in\mathbb Z\}$ & \multirow{2}*{Cannot} & \multirow{2}*{Always} \\
    & ~$\mu=0$ or $\varphi_1 \in\{k\pi+\pi/2,k\in\mathbb Z\}$ or $\varphi_1 \in\{k\pi,k\in\mathbb Z\}$~ & & \\\hline
    $R_{IV}$ & $\varphi_1 \in\{k\pi,k\in\mathbb Z\}$ and $\chi \in\{k\pi/4,k\in\mathbb Z\}$ & Always & Cannot\\
\hline\hline
\end{tabular}
    \caption{Conditions for Yang-Baxter gates defined in Eqs. (\ref{eq:R_1}), (\ref{eq:R_2}), and (\ref{eq:R_3}) to be the Clifford gate, the matchgate, and the dual unitary gate. The conditions for $R_{II,1,2,3}$ are the same as the conditions for $R_{II,1,2,3}$, therefore omitted in the table. }
    \label{tab:YBG_conditions}
\end{table*}

Based on the gate decomposition of $R_I$ shown in Fig. \ref{fig:R_I_decomposition}, the necessary and sufficient condition for $R_I$ to be the Clifford gate is $\omega\in\{k\pi,k\in\mathbb Z\}$, and $a_{I,1},c_{I,1},\varphi_{I,2},\varphi_{I,3}\in\{k\pi/2,k\in\mathbb Z\}$. Specifically, the $R_{I,1}$ is a Clifford gate if $\mu = 0$, or $\varphi\in \{k\pi,k\in\mathbb Z\}$. The former case gives the trivial identity gate, while the latter is locally equivalent to the SWAP gate. For $R_{I,2}$ and $R_{I,3}$, when $\mu = 0$ or $\varphi\in \{k\pi,k\in\mathbb Z\}$, we have $\varphi_{I,2}$ equal to $0$ or $\pi$, which gives the locally equivalent SWAP gate (therefore to be the Clifford gate). Additionally, if the spectral parameter $\mu$ satisfies
\begin{equation}
\label{eq:Clifford_condition_R_I}
    \tanh\left(\frac{\mu}{2}\right) = \pm\tan\left(\frac{\varphi}{2}\right),
\end{equation}
we can also have $R_{I,2}$ as the Clifford gate, which is locally equivalent to the iSWAP gate. For $R_{I,3}$, it is also true if
\begin{equation}
\label{eq:Clifford_condition_R_II}
    \tanh\left(\frac{\mu}{2}\right) = \pm\cot\left(\frac{\varphi}{2}\right),
\end{equation}
is satisfied, which is the correspondence $\varphi\rightarrow \pi/2-\varphi$ from Eq. (\ref{eq:Clifford_condition_R_I}). 

The analysis on $R_{II}$ to be the Clifford gate is the same as $R_{I}$ due to the relation (\ref{eq:R_I_to_R_II}). For the case $R_{III}$, we see that there are $T$ gates in the decompositions. See Fig. \ref{fig:R_III_decomposition}. Therefore, the necessary and sufficient condition for $R_{III}$ to be the Clifford gate becomes $\varphi_2\in\{k\pi+\pi/2,k\in\mathbb Z\}$, and $a_{III,1},c_{III,1},\varphi_{III,2},\varphi_{III,3}\in\{k\pi/2,k\in\mathbb Z\}$. Specifically, the condition $a_{III,1},c_{III,1}\in\{k\pi/2,k\in\mathbb Z\}$ (for $R_{III,1}$ to be the Clifford gate) is satisfied if $\mu = 0$ or $\varphi_1\in \{k\pi/2,k\in\mathbb Z\}$. It is similar with $R_{I,1}$. When $\mu=0$ or $\varphi_1 \in\{k\pi+\pi/2,k\in\mathbb Z\}$, we have the Clifford $R_{III,2}$, which is locally equivalent to iSWAP gate; when $\varphi_1 \in\{k\pi,k\in\mathbb Z\}$, we have the Clifford $R_{III,2}$, which is locally equivalent to SWAP gate. Similar results are applied to $R_{III,3}$. The Yang-Baxter gate $R_{IV}$ is Clifford gate if $\varphi_1 \in\{k\pi,k\in\mathbb Z\}$ and $\chi \in\{k\pi/4,k\in\mathbb Z\}$, which is locally equivalent to CNOT. We summarize the results in Table \ref{tab:YBG_conditions}.

The matchgate must have a vanishing nonlocal parameter \cite{brodExtendingMatchgatesUniversal2011}. We find that the $R_{I,1}$ gate is a nontrivial matchgate (not proportional to the identity operator) if $\varphi = \pi/2$. The $R_{I,2}$ gate is a matchgate if it is locally equivalent to the iSWAP gate where the spectral parameter satisfies the Eq. (\ref{eq:Clifford_condition_R_I}). The $R_{I,3}$ gate has the similar results. Note that the Pauli transformation (\ref{eq:R_I_to_R_II}) preserves the matchgate. Therefore $R_{II}$ is the matchgate if $R_I$ is the matchgate with the same parameters. In the case of $R_{III}$, it can not be a nontrivial matchgate, similar to $B_{III}$. While $R_{IV}$ is always a matchgate independent of the spectral parameter. 

The dual unitary two-qubit gates are on the edge $A_2A_3$ of the two-qubit tetrahedron \cite{bertiniExactCorrelationFunctions2019}. Therefore, $R_{l,2}$ and $R_{l,3}$ with $l\in\{I,II,III\}$ are always dual unitary independent on the spectral parameter. However, $R_{IV}$ can not be the dual unitary gate, since it is on the edge $OA_1$. See Fig. \ref{fig:geo_YBG}. When $R_{l,1}$ with $l\in\{I,II,III\}$ reduces to the corresponding braid gate (by taking the spectral parameter $x=0$), it is on the edge $A_2A_3$, therefore are dual unitary.

\subsection{\label{subsec:YBG_entangling}Entangling powers}

\begin{figure*}[t]
\centerline{\includegraphics[width=\textwidth]{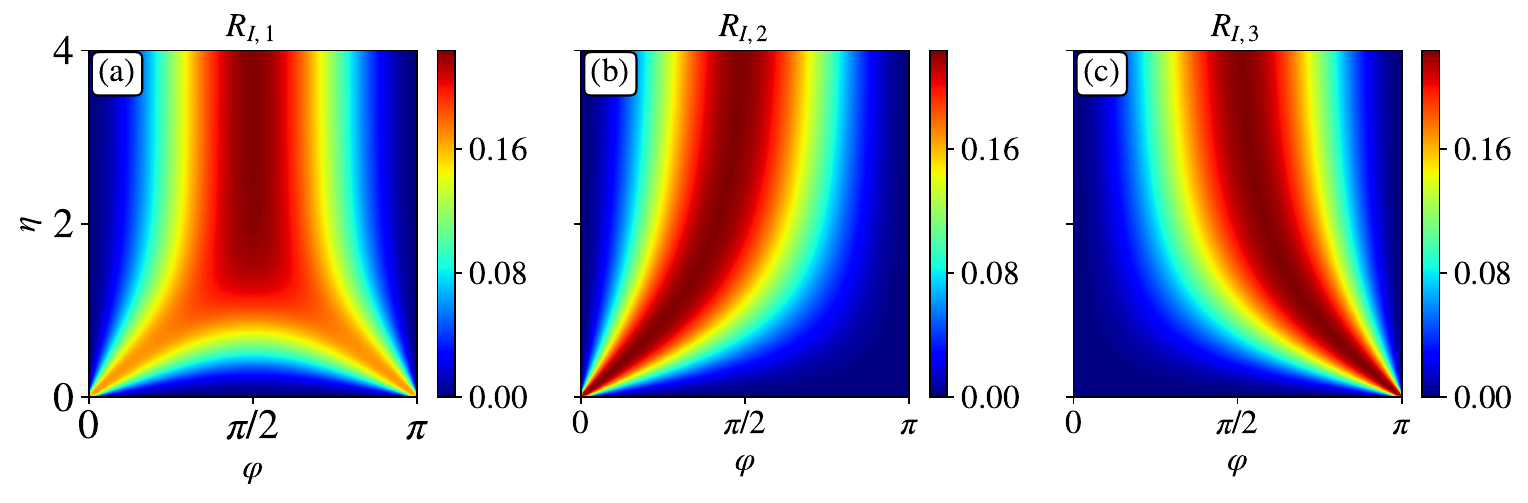}}
\caption{Entangling powers of Yang-Baxter gates $R_{I}$ obtained from braid gate $B_I$. Yang-Baxter gates $R_{I,1}$, $R_{I,2}$, and $R_{I,3}$ are defined in Eqs. (\ref{eq:R_1}), (\ref{eq:R_2}), and (\ref{eq:R_3}) respectively. Here $\mu = \ln(x)$ is the spectral parameter, and $\varphi = (\varphi_2+\varphi_3)/2-\varphi_1$ is the braid gate parameter. }
\label{fig:YBG_entangling_power}
\end{figure*}

The entangling power of the two-qubit gate defined in Eq. (\ref{def:entangling_power}) quantifies the ability to generate entanglement from the product state. It can be viewed as the interacting strength between qubits in the brick-wall circuits \cite{hahn2024absence}. Given the nonlocal parameters of $R_I$, we numerically calculate its entangling powers in terms of the braid gate parameter $\varphi$ and the spectral parameter $\mu$. See Fig. \ref{fig:YBG_entangling_power}. Specifically, when $\mu=0$ or $\varphi\in \{k\pi,k\in\mathbb Z\}$, $R_{I,1,2,3}$ has the zero entangling power, corresponding to the trivial two-qubit gate or SWAP gate. When $\mu\rightarrow\infty$ and $\varphi\in \{k\pi+\pi/2,k\in\mathbb Z\}$, $R_{I,1,2,3}$ has the maximal entangling power, corresponding to the iSWAP gate. Note that $\mu \leftrightarrow -\mu$ does not change the magnitude of the nonlocal parameters and the entangling power, therefore we only draw $\mu\geq0$ in Fig. \ref{fig:YBG_entangling_power}.

The entangling powers of $R_{II,1,2,3}$ are exactly same as $R_{I,1,2,3}$ due to the relation of Eq. (\ref{eq:R_I_to_R_II}). The entangling powers of $R_{III,1}$ is qualitatively same as $R_{I,1}$ with the correspondence $2\varphi_1 \leftrightarrow \varphi$ and $2\mu\leftrightarrow \mu$. The Yang-Baxter gate $R_{IV}$ has the entangling power
\begin{equation}
    e_p(R_{IV}) = \frac 2 9 \sin^2(2\chi),
\end{equation}
which has the maximal at $\chi = \pi/4$, corresponding to $B_{IV}$. Note that the Yang-Baxter gates have the maximal entangling power corresponding to the point $A_2$ with the nonlocal parameters $[\pi/2,\pi/2,0]$ (iSWAP gate), or the middle of $OA_1$ with the nonlocal parameters $[\pi/2,0,0]$ (CNOT gate).

Both the phase parameter $\varphi$ and the spectral parameter $\eta$ determine the entangling power of the Yang-Baxter gates. Recall that the spectral parameter is related to the Trotter step of the brick-wall circuit. If we aim to construct states with high entanglement, then the Trotter step should be large with the fixed phase parameter $\varphi = \pi/2$. If we aim to simulate the dynamics of the integrable system with high fidelity, then the Trotter step should be small while avoiding the phase parameter giving large entangling power. This is because the entangled state is more fragile to the noises than the product state.

\section{\label{sec:conclusion}Conclusions}

In this work, we have studied the geometric representations of braid and Yang-Baxter gates, where the Yang-Baxter gates are obtained from the Yang-Baxterization of braid gates. Braid and Yang-Baxter gates on the same point of the two-qubit tetrahedron can be transformed from each other through single-qubit gates. In the two-qubit tetrahedron representation, we find that the Yang-Baxter gates can only exist on the edges $A_2A_3$ and $OA_1$, and on the faces $OA_2A_3$ and $A_1A_2A_3$. Based on their geometric representations, we give their gate decompositions in terms of the universal gate set $\{R_z(\theta),H,\text{CNOT}\}$ with the minimal number of CNOT. The braid and Yang-Baxter gates have the maximal entangling power if they are locally equivalent to iSWAP or CNOT. We also find that the Yang-Baxter gates can be the Clifford gate, the matchgate, and the dual unitary gate for specific parameters. It would be interesting to explore the dynamics of the brick-wall circuits composed of such special Yang-Baxter gates. For example, the Yang-Baxter matchgate corresponds to the free-fermion integrable model; the dual unitary Yang-Baxter gate may give special constraints on the speed of information spreading in the integrable circuits.


Certainly, we do not cover all the Yang-Baxter gates in our study. Finding all (parameter dependent) solutions of the Yang-Baxter equation, even for the two-qubit representation, is challenging \cite{garkun2024new}. It would be interesting to check whether other Yang-Baxter gates can exist in other regions of the two-qubit tetrahedron. Our study can also be generalized to the unitary solution of the colored Yang-Baxter equation, also called the multi-parametric Yang–Baxter equation or the Yang-Baxter equation with non-additive spectral parameters \cite{zhang2020new,padmanabhan2024integrability}. Most studies on the integrable circuit are limited to the one-dimensional qubit array. It is natural to study the unitary solutions of the tetrahedron equation, which is the multidimensional generalization of the Yang-Baxter equation \cite{zamolodchikov1981tetrahedron,padmanabhan2024solving,sinha2024toffoli}. In other words, the integrable circuits might be able to be generalized to the two-dimensional qubit array. We leave it for our future study.

\section*{Data availability statement}

All data needed to evaluate the conclusions in the paper are present in the paper. All data that support the findings of this study are included within the article (and any supplementary files).

\ack

The work of K.Z. was supported by the National Natural Science Foundation of China under Grant Nos. 12305028 and 12247103, and the Youth Innovation Team of Shaanxi Universities. The work of K.H. was supported by the National Natural Science Foundation of China (Grant Nos. 12275214, 12247103, and 12047502), the Natural Science Basic Research Program of Shaanxi Province Grant Nos. 2021JCW-19 and 2019JQ-107, and Shaanxi Key Laboratory for Theoretical Physics Frontiers in China. K.Y. and V.K. are funded by the U.S. Department of Energy, Office of Science, National Quantum Information Science Research Centers, Co-Design Center for Quantum Advantage under Contract No. DE-SC0012704. 

\appendix

\section{\label{App:A} Proof of Theorem \ref{theorem_real_x}}

\begin{proof}
    Through Yang-Baxterization of the braid matrix with two distinct eigenvalues, we have
    \begin{align}
        R(x) = & \left(x+\frac{\lambda_1}{\lambda_2}\right)P_1 + \left(1+x\frac{\lambda_1}{\lambda_2}\right)P_2 \nonumber \\
        = & \frac{1}{\lambda_2}\left(B+x\lambda_1\lambda_2B^{-1}\right).
    \end{align}
    Since we assume that $B$ is unitary, therefore $|\lambda_1| = |\lambda_2| = 1$ and $B^{-1} = B^\dag$. The unitary condition is given by
    \begin{align}
        R(x)R^\dag(x) = 1\!\!1 + |x|^2 + x\lambda_1\lambda_2{B^\dag}^2 + x^*\lambda_1^*\lambda_2^*B^2,
    \end{align}
    with the complex conjugate $*$. Through the spectral decomposition of $B$, we have
    \begin{align}
        {B^\dag}^2 = {\lambda^*_1}^2P_1 + {\lambda^*_2}^2P_2
        = {\lambda^*_1}^2 + {\lambda^*_2}^2 - {\lambda^*_1}^2{\lambda^*_2}^2B^2,
    \end{align}
    which gives
    \begin{equation}
        R(x)R^\dag(x) = 1\!\!1 + |x|^2 + x\left(\lambda^*_1\lambda_2+\lambda^*_2\lambda_1\right) + \left(x^* -x\right)\lambda^*_1\lambda^*_2B^2.
    \end{equation}
    If $B^2\neq1\!\!1$, the spectral parameter must be a real number for unitary $R(x)$.
\end{proof}

\quad 

\section{\label{App:B} matrix expression of Yang-Baxter gates}

Consider the Yang-Baxter equations with the additive spectral parameter 
\begin{equation}
    \mu = \ln x,
\end{equation}
and $x>0$. The braid gates $B_I$ and $B_{II}$ give the Yang-Baxter gates 
\begin{subequations}
    \begin{equation}
    \small
        R_{I,1}(\mu)  \cong  \begin{pmatrix}     
            1 & 0 & 0 & 0 \\
            0 & \dfrac{\sin\varphi}{\sin(\varphi-i\mu)} & \dfrac{-ie^{i\omega}\sinh\mu}{\sin(\varphi-i\mu)} & 0 \\
            0 & \dfrac{-ie^{-i\omega}\sinh\mu}{\sin(\varphi-i\mu)} & \dfrac{\sin\varphi}{\sin(\varphi-i\mu)} & 0 \\
            0 & 0 & 0 & 1 \\
            \end{pmatrix},
    \end{equation}
    \begin{equation}
    \small
        R_{I,2}(\mu)  \cong \frac{1}{\sqrt{\Delta_{I,2}}}\begin{pmatrix}
            \sinh\left(\frac{1}{2}(\mu+i\varphi)\right) & 0 & 0 & 0 \\
            0 & 0 & e^{i\omega}\sinh\left(\frac{1}{2}(\mu-i\varphi)\right) & 0 \\
            0 & e^{-i\omega}\sinh\left(\frac{1}{2}(\mu-i\varphi)\right) & 0 & 0 \\
            0 & 0 & 0 & \sinh\left(\frac{1}{2}(\mu+i\varphi)\right) \\
            \end{pmatrix},
    \end{equation}
    \begin{equation}
    \small
        R_{I,3}(\mu)  \cong \frac{1}{\sqrt{\Delta_{I,3}}}\begin{pmatrix}
            \cosh\left(\frac{1}{2}(\mu+i\varphi)\right) & 0 & 0 & 0 \\
            0 & 0 & e^{i\omega}\cosh\left(\frac{1}{2}(\mu-i\varphi)\right) & 0 \\
            0 & e^{-i\omega}\cosh\left(\frac{1}{2}(\mu-i\varphi)\right) & 0 & 0 \\
            0 & 0 & 0 & \cosh\left(\frac{1}{2}(\mu+i\varphi)\right) \\
            \end{pmatrix},
    \end{equation}
    \begin{equation}
    \small
        R_{II,1}(\mu) \cong \begin{pmatrix}   
            \dfrac{\sin\varphi}{\sin(\varphi-i\mu)} & 0 & 0 & \dfrac{-ie^{i\omega}\sinh\mu}{\sin(\varphi-i\mu)} \\
            0 & 1 & 0 & 0 \\
            0 & 0 & 1 & 0 \\
            \dfrac{-ie^{-i\omega}\sinh\mu}{\sin(\varphi-i\mu)} & 0 & 0 & \dfrac{\sin\varphi}{\sin(\varphi-i\mu)} \\
            \end{pmatrix},
    \end{equation}
    \begin{equation}
    \small
        R_{II,2}(\mu)  \cong \frac{1}{\sqrt{\Delta_{II,2}}}\begin{pmatrix}
            0 & 0 & 0 & e^{i\omega}\sinh\left(\frac{1}{2}(\mu-i\varphi)\right) \\
            0 & \sinh\left(\frac{1}{2}(\mu+i\varphi)\right) & 0 & 0 \\
            0 & 0 & \sinh\left(\frac{1}{2}(\mu+i\varphi)\right) & 0 \\
            e^{-i\omega}\sinh\left(\frac{1}{2}(\mu-i\varphi)\right) & 0 &  0 & 0 \\
            \end{pmatrix}, 
    \end{equation}
    \begin{equation}
    \small
        R_{II,3}(\mu) \cong \frac{1}{\sqrt{\Delta_{II,3}}}\begin{pmatrix}
            0 & 0 & 0 & e^{i\omega}\cosh\left(\frac{1}{2}(\mu-i\varphi)\right) \\
            0 & \cosh\left(\frac{1}{2}(\mu+i\varphi)\right) & 0 & 0 \\
            0 & 0 & \cosh\left(\frac{1}{2}(\mu+i\varphi)\right) & 0 \\
            e^{-i\omega}\cosh\left(\frac{1}{2}(\mu-i\varphi)\right) & 0 &  0 & 0 \\
            \end{pmatrix},
    \end{equation}
\end{subequations}
with the defined parameters
\begin{subequations}
    \begin{align}
        & \varphi =  \frac 1 2(\varphi_2+\varphi_3)-\varphi_1, \\
        & \omega =  \frac 1 2(\varphi_2-\varphi_3), 
    \end{align}
\end{subequations}
and the normalizations 
\begin{subequations}
\begin{align}
    & \Delta_{I,2} = \Delta_{II,2} = \sin^2\left(\frac\phi 2\right) + \sinh^2\left(\frac\mu 2\right), \\
    & \Delta_{I,3} = \Delta_{II,3} = \cos^2\left(\frac\phi 2\right) + \sinh^2\left(\frac\mu 2\right).
\end{align}
\end{subequations}

The three kinds of Yang-Baxter gates obtained from $B_{III}$ have the matrix expression
\begin{subequations}
    \begin{equation}
    \small
        R_{III,1}(\mu) \cong \begin{pmatrix}
            \dfrac{\cosh\mu\cos\varphi_1}{\cosh(\mu+i\varphi_1)} & 0 & 0 & -e^{i\varphi_2}\dfrac{\sinh\mu\sin\varphi_1}{\cosh(\mu+i\varphi_1)} \\
            0 & i\dfrac{\cosh\mu\sin\varphi_1}{\sinh(\mu+i\varphi_1)} & -\dfrac{\sinh\mu\cos\varphi_1}{\sinh(\mu+i\varphi_1)} & 0 \\
            0 & -\dfrac{\sinh\mu\cos\varphi_1}{\sinh(\mu+i\varphi_1)} & i\dfrac{\cosh\mu\sin\varphi_1}{\sinh(\mu+i\varphi_1)} & 0 \\
            e^{-i\varphi_2}\dfrac{\sinh\mu\sin\varphi_1}{\cosh(\mu+i\varphi_1)} & 0 & 0 & \dfrac{\cosh\mu\cos\varphi_1}{\cosh(\mu+i\varphi_1)}
            \end{pmatrix},
    \end{equation}
    \begin{equation}
    \small
        R_{III,2}(\mu) \cong \frac{1}{\sqrt{\Delta_{III,2}}}\begin{pmatrix}
            \sinh\mu\cos\varphi_1  & 0 & 0 & e^{i\varphi_2}\cosh\mu\sin\varphi_1  \\
            0 & i\cosh\mu\sin\varphi_1  & -\sinh\mu\cos\varphi_1 & 0 \\
            0 & -\sinh\mu\cos\varphi_1 & i\cosh\mu\sin\varphi_1 & 0 \\
            -e^{-i\varphi_2}\cosh\mu\sin\varphi_1  & 0 & 0 & \sinh\mu\cos\varphi_1 \\
            \end{pmatrix},
    \end{equation}
    \begin{equation}
    \small
        R_{III,3}(\mu) \cong \frac{1}{\sqrt{\Delta_{III,3}}}\begin{pmatrix}
            \cosh\mu\cos\varphi_1  & 0 & 0 & -e^{i\varphi_2}\sinh\mu\sin\varphi_1  \\
            0 & i\sinh\mu\sin\varphi_1  & -\cosh\mu\cos\varphi_1 & 0 \\
            0 & -\cosh\mu\cos\varphi_1 & i\sinh\mu\sin\varphi_1 & 0 \\
            e^{-i\varphi_2}\sinh\mu\sin\varphi_1  & 0 & 0 & \cosh\mu\cos\varphi_1 \\
            \end{pmatrix},
    \end{equation}
\end{subequations}
with the normalizations
\begin{subequations}
    \begin{align}
        \Delta_{III,2} = & \sinh^2\mu\cos^2\varphi_1+\cosh^2\mu\sin^2\varphi_1,\\
        \Delta_{III,3} = & \cosh^2\mu\cos^2\varphi_1+\sinh^2\mu\sin^2\varphi_1. 
    \end{align}
\end{subequations}


\providecommand{\noopsort}[1]{}\providecommand{\singleletter}[1]{#1}%
\providecommand{\newblock}{}

\end{document}